\input psfig.sty
\documentclass{./aa}

\usepackage{graphicx}

\begin{document}

\title{HST observations rule out the association between Cir X-1 and SNR G321.9-0.3\thanks{Based
on observations with the NASA/ESA Hubble Space Telescope,
obtained at the Space Telescope Science Institute, which is
operated by AURA, Inc.  under contract No NAS 5-26555}}

\author{R.P. Mignani\inst{1} \and A. De Luca\inst{2,3} \and P.A.
Caraveo\inst{2}
\and I.F. Mirabel\inst{4,5} }

\offprints{rmignani@eso.org}

\institute{ESO, Karl Schwarzschild Str.  2, D-85748 Garching bei
M\"unchen,  Germany  \and Istituto  di  Fisica  Cosmica  del CNR  ``G.
Occhialini", Via Bassini 15, I-20133 Milan, Italy \and
Universit\'a di Milano Bicocca, Dipartimento di Fisica, Piazza della
Scienza 3, I-20126 Milano, Italy  \and Service
d'Astrophysique / CEA, CE-Saclay, 91191
Gif/Yvette, France \and Instituto de Astronom\'\i a y F\'\i sica del
Espacio/CONICET. cc5, 1428 Bs As, Argentina}

\date{Received / Accepted }

\titlerunning{Cir X-1}

\abstract{Cir X-1 is one of the most intriguing galactic X-ray sources. It
is a $\sim 16.6$ days  variable X/radio source, a type-I X-ray burster
and  a QPO  emitter.  In  spite  of an  uncertain optical  counterpart
classification, all  these properties identify the source  as an LMXB.
The  morphology  of the  surrounding  radio  nebula  has suggested  an
association  with  the   nearby  ($\sim  25$~arcmin)  SNR  G321.9-0.3,
implying  that  Cir  X-1  is  a runaway  binary  originated  from  the
supernova  explosion $\approx  10^5$ years  ago.  To  investigate this
hypothesis, we carried out a  proper motion measurement of the Cir X-1
$m  \sim 19$  optical counterpart  using a  set of  HST/WFC  and WFPC2
observations taken $\sim$  8.6 years apart.  We obtained  a $3 \sigma$
upper limit of $\approx$ 5 mas yr $^{-1}$ on the source proper motion.
Since the  runaway hypothesis would  have implied a proper  motion due
North ranging between 15 and 75 mas yr$^{-1}$, depending on the actual
age  of the  SNR, our  result definitively  rules out  the association
between  Cir X-1  and SNR  G321.9-0.3.   }

\maketitle

\section{Introduction.}

Cir X-1 is a strong and persistently bright galactic X-ray source with
a flux  modulated over a ~16.6  days cycle (Kaluzienski  et al.  1976,
Bignami  et al.  1977),  which is commonly believed  to coincide
with the orbital period of a binary system in a highly eccentric orbit
(Murdin et  al.  1980).  Type-I  X-ray bursts have been  discovered by
EXOSAT  (Tennant, Fabian  and  Shafer 1986a,b),  as  well as  high/low
frequency QPO (Tennant 1987,1988)  whose properties suggested that Cir
X-1 might represent an ideal link between the classes of the canonical
{\em Z} and  {\em atoll}-type sources (Oosterbroek et  al.  1995).  In
addition, long-term  variations in the X-ray flux  have been observed,
probably due  to a  partial obscuration of  the central source  by the
precession   of  a   warped   accretion  disk   (Oosterbroek  et   al.
1995). Recently, X-ray spectral evolution of Cir X-1 along the orbital
cycle has been reported by Iaria  et al. (2001).  \\ While the general
X-ray properties of Cir X-1  (e.g.  type-I bursts, QPOs) point towards
a LMXB, its nature is  still unclear.  Although an optical counterpart
has been  identified (Moneti 1992; Mignani, Caraveo  \& Bignami 1997),
no  reliable spectral classification  has been  possible so  far.  The
spectrum is dominated by  a broad, asymmetric, H$\alpha$ line (Duncan,
Stewart  \& Haynes  1993; Mignani,  Caraveo \&  Bignami 1997)  with no
prominent stellar  emission/absorption features (Johnston,  Fender and
Wu  1999). \\ The  radio counterpart  of Cir  X-1 was  identified soon
after  the  X-ray detection  (Clark  et al.   1975)  and  found to  be
variable with  the same  16.6 day period  (Haynes et al.   1978).  The
radio picture features a point source which is embedded in a $\simeq 5
\times 10$ arcmin$^2$ nebula and is apparently  the origin of elongated
structures, extending on both arcmin  (Stewart et al. 1993) and arcsec
scales  (Fender  et al.   1998).    Such a morphology   has  been
interpreted as  a jet-powered synchrotron  nebula, making Cir  X-1 the
second galactic  X-ray binary, besides  SS433, which shows  both radio
jets  and a  radio  synchrotron nebula.   \\  The shape  of the  radio
nebula, protruding southward and pointing  to the center of the nearby
($\sim 25$~arcmin) radio SNR  G321.9-0.3, is reminiscent of a residual
wake left  behind as  the point  source moves away  from the  SNR (see
Fig.1, reproduced from Fig.1 of Stewart et al.  1993).  This suggested
a  tantalizing scenario where Cir X-1 is a runaway binary system,
with  the  neutron star  member  ejected  by  the supernova  explosion
responsible  for SNR  G321.9-0.3  (Stewart et  al.   1993) between  $2
\times 10^{4}$ and  $10^{5}$ yrs ago (Clark, Caswell  and Green 1975).
Such  a scenario  is consistent,  within the  uncertainties,  with the
estimated  distances  of both  Cir  X-1  and of  the  SNR.   From H  I
absorption measurements (Goss
\& Mebold 1977), the distance to Cir X-1 is anywhere between 6.7 and 8
kpc,  while the  distance to  the  SNR ranges  from 5.5  kpc (Case  \&
Battacharya 1998)  to 6.5  kpc (Stewart et  al.  1993).  If  the above
scenario is  correct, for  a SNR  age $T$ (yrs),  Cir X-1  should move
Northward in the plane of the sky  with a proper motion $\mu \sim 15 ~
(T/10^{5} yrs)^{-1}$ mas yr$^{-1}$. This value would imply a tranverse
velocity with respect  to the center of the  SNR of $v \sim 5  ~ \mu ~
d_{kpc}$ km  s$^{-1}$, where  $\mu$ is in  units of mas  yr$^{-1}$ and
$d_{kpc}$  is  the  SNR  distance  in  units  of  kpc.   \\  From  the
combination  of  all the  possible  values of  both  the  SNR age  and
distance, we would thus expect Cir X-1 to move Northward with a proper
motion ranging  between 15  and 75 mas  yr$^{-1}$, corresponding  to a
transverse velocity between 400 and  2400 km s$^{-1}$.\\ Thus, the Cir
X-1 angular displacement is a simple, yet  powerful, tool to prove    or     disprove    its    association    with    SNR
G321.9-0.3. Measuring  a Northward angular displacement $  \ge$ 15 mas
yr  $^{-1}$ would secure  the association,  while the  lack of  such a
displacement, or  a displacement in a different  direction, would rule
it out. In principle, both  the optical and radio counterpart could be
used  to gauge  the  precise  position of  the  source.  However,  the
dramatic fading of  the radio point source (Stewart  et al. 1991) 
hampers a measurement  in the radio band and leaves  the burden of the
proof to  astrometry of the  optical counterpart.\\ The  sharpness of
view of  HST makes  it the  ideal instrument to  settle such  a proper
motion issue.  Thus, we  have taken advantage  of an  HST/WFC archival
image of the field as well as of recently acquired WFPC2 ones to carry
out  the  proper  motion   measurement.   The  observations  and  data
reduction are described  in Section 2 while the  results are presented
and discussed in Sections 3 and 4.

\section{Observations.}

\subsection{The data sets}

The  first image  of the  Cir  X-1 field  was obtained   with  the
pre-repair  HST on  October $20^{th}$  1992 as  a  target acquisition
image for spectroscopy observations with the Faint Object Spectrograph
(Mignani et al.   1997).  The observation was performed  with the Wide
Field  Camera  (WFC) (0\farcs1/pixel)  through  the  wide band  filter
$785LP$ ``I-band filter'' ($\lambda=9703$ \AA; $\Delta \lambda=1669.5$
\AA),  where the  source  is brighter  ($m_{785LP}  \simeq 18$).   The
exposure time was  300 s.  A new observation of  Cir X-1 was performed
on  May $19^{th}$  2001 with  the Wide  Field and  Planetary  Camera 2
(WFPC2), as  a snapshot  programme.  In order  to achieve  the highest
positional accuracy,  the target was  positioned at the center  of the
Planetary Camera (PC) chip ($0\farcs045$/pixel).   For consistency
with   the  first observation,  the  $785LP$ filter  was used.  Three
observations of  300 s each were  obtained to allow for  a better 
cosmic  ray filtering.   After  the standard  HST pipeline  reduction
(debiassing,  dark removal,  flatfielding) the  2001 images  have been
combined with a median filter  and stacked (Fig.  2).  Residual cosmic
ray hits were filtered out using specific IRAF/STSDAS routines.

\begin{figure}
\centerline{\hbox{\psfig{figure=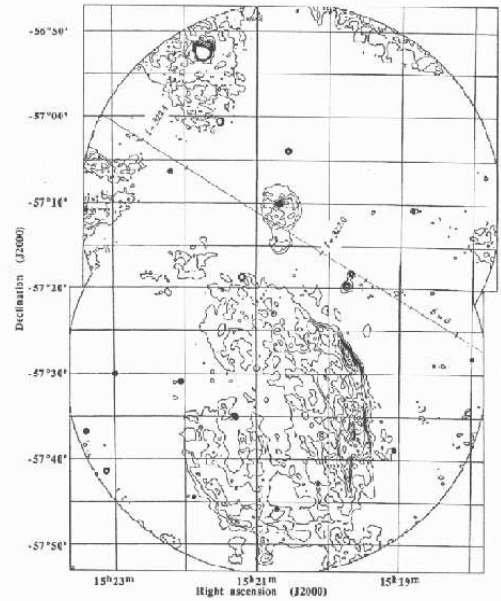,height=11cm,clip=}}}
\caption{ATCA radio map of the Cir X-1 region, adapted from Fig.1 of
Stewart et al. (1993). Cir X-1 is at the center of the compact radio
nebula recognizable near the center of the field, with SNR G321.9-0.3
due south of it.}
\end{figure}

\subsection{The data analysis}

The  cleaned  images  are  the  starting  point  for  our  astrometric
analysis.   Following  a robust  method  successfully  applied  in  several
previous astrometric works (see e.g.  Caraveo et al.  1996; Caraveo \&
Mignani 1999; De Luca et al.  2000,2001; Mignani et al. 2000,2001; Caraveo
et al.  2001), this relies on an accurate image superposition which is
obtained by optimizing a  linear coordinate transformation between two
reference grids determined by a set of common reference objects.

\begin{figure}
\centerline{\hbox{\psfig{figure=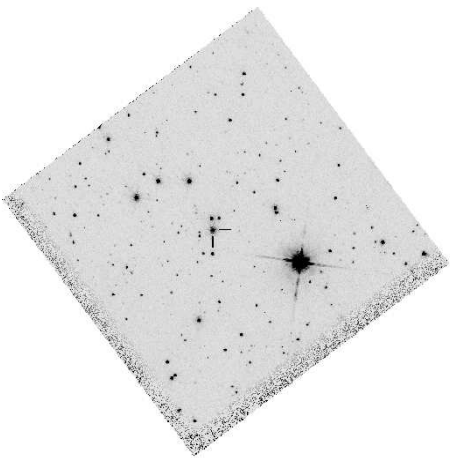,height=9cm,clip=}}}
\caption{WFPC2 image of the Cir X-1 field taken in 2001 with the WFPC2 
through the F785LP filter. Only the PC field is shown. North is up,
East is left. The two ticks  mark the position of the companion star of the X-ray source.}
\end{figure}

In order  to obtain an accurate position  determination, the reference
objects have  to be carefully selected  i.e. they must be  (i) not too
close to the CCD edges, (ii) not extended, (iii) not saturated but not
too faint.  In order to  avoid problems with the inter-chip astrometry
(see Caraveo \&  Mignani 1999), the selecton of  the reference objects
was limited to the chip where the target star was imaged, i.e. WFC \#2
in the 1992 observation and PC  in the 2001 one. In the rather crowded
field  of Cir  X-1,  46  good reference  sources,  matching the  above
constraints, could be found.  Their positions were computed with a 2-D
gaussian  fitting of  their intensity  profiles using  centering boxes
optimized  in order  to obtain  stable values  for both  the centroids
coordinates and  their uncertainties (see De  Luca et al.   2001 for a
detailed discussion of both statistical and systematic errors involved
in  the procedure).   The reference  objects were  positioned  with an
accuracy  between 0.05 and 0.08 pixel (i.e.  between 5 and 8 mas)
in the WFC  image and  between 0.05 and  0.1 pixel (i.e.  between
2.3 and  4.5 mas) in the PC  one, depending on the  object's S/N.  The
position of  the target  star was  computed in the  same way,  with an
accuracy of about 0.07 pixel (7  mas) and of about 0.06 pixel ($\sim$3
mas) in  the WFC and in the  PC image, respectively. We  note that the
uncertainties on the  centroids in both the WFC and  the PC images are
somewhat higher than  the ones measured in other  fields (see e.g.  De
Luca et  al. 2000).  This can be  ascribed, at  least in part,  to the
peculiar PSF  in the long-pass  F785LP filter, characterized  by broad
and  relatively bright  wings. This  is  the same  effect observed  in
filter F1042M  (it is known as  the ``F1042M PSF anomaly'')  and it is
due to the CCD detector becoming transparent at longer wavelengths, so
that light is reflected and scattered by the back of the CCD producing
a  defocused   halo  (see   WFPC2  Instrument  Handbook   for  further
details). In addition, for the 2001 snapshot observations the analysis
of the  spacecraft attitude showed a  jitter of the  telescope about 2
times larger than average ($\approx$ 5 mas rms  compared to 3 mas
rms) and distributed along a preferential direction. Moreover, we note
that  the  2001  observations  were  performed  during  the  day/night
transition,  which  occurred   during  the  first  exposure.   Thermal
effects, such  as ``breathing'', could thus be  present, determining a
slight defocussing of the camera.  All these effects clearly influence
the PSF, thus affecting the centering accuracy. \\ The derived sets of
coordinates  were  then  corrected  for known  systematics,  like  the
"34$^{th}$ row defect" of the PC (Anderson
\& King 1999), and for the effects of the  distortions of the WFC
and the  PC.  While for the  WFC only the  geometric transformation of
Gilmozzi et al.   (1992) is available, for the WFPC2  we have used the
most recent one provided by  Casertano \& Wiggs (2000), which accounts
for the evolution of the   distortion map of the WFPC2 due to the
drift of the instrument in the Optical Telescope Assembly focal plane.
\\ The superposition of the frames was performed using the PC image as
reference.  First,  the PC  image was aligned  in Right  Ascension and
Declination,  according to the  telescope roll  angle.  Then,  the WFC
reference  grid was  registered  on the  PC  one by  fitting a  linear
coordinate transformation, accounting for 2 independent scale factors,
2 translation factors and a  rotation angle.  Following Caraveo et al.
(2001), we checked for systematics due, e.g., to unknown displacements
of the reference objects by repeating the frame superposition 46 times
excluding, in turn, one of our reference stars.  Four objects yielding
relatively high deviations (between 0.5 and 1 PC pixel per coordinate)
were  thus identified  and rejected.   Different  centering algorithms
(e.g.   separate  gaussian  fitting  on  the  $X$  and  $Y$  intensity
distributions)  and  coordinate   transformation  routines  were  also
tested.     In   all    cases   we    obtained    statistically   
indistinguishable results, showing that  our procedure is correct and
free of  systematics.  \\ The  final frame registration yielded  a rms
values on the residuals of  the reference objects' coordinates of 0.33
PC pixel (15  mas) in Right Ascension and 0.25 PC  pixel (11.4 mas) in
Declination.   Such rms values  are a  factor of  4 higher  than those
obtained by De  Luca et al.  (2000a) for  a WFC-to-PC superposition in
the  Vela  pulsar  field.   While  the  higher  uncertainties  in  the
centroids  determination addressed  above play  a role,  they  can not
account  entirely  for the  larger  residuals  obtained  in the  image
registration.
This effect  might be related to  the likely dependance  on the filter
wavelength of  the transformation  coefficients used in  the geometric
correction.    Unfortunately,   since   such   dependance   has   been
characterized for  the WFPC2  (Trauger et al.   1995) but not  for the
WFC, the  accuracy of  the image registration  might be affected  by a
non-optimized geometric correction.

\begin{figure}
\centerline{\hbox{\psfig{figure=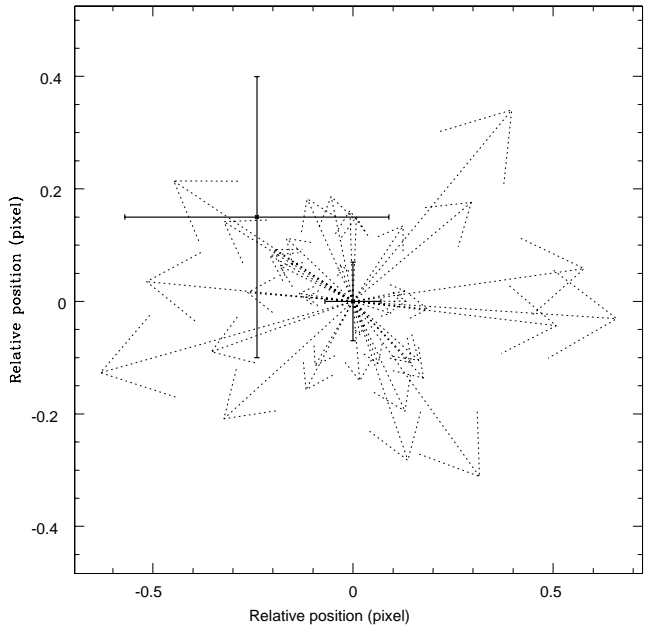,height=9cm,clip=}}}
\caption{Relative positions (crosses) of the Cir X-1 companion
star measured at the epoch of our two observation. The error bars
for the 2001 position (defining the 0,0 in the plot) account only
for the centering uncertainty, while for the  1992  position  they
account  also for  the  frame  registration uncertainty. We
overplotted as dashed arrows the residuals  on the reference
stars  coordinates after  the image superposition.  Their rms
value represents the frame  registration uncertainty (see text).
The displacement of Cir X-1 is comparable with the residuals of
the reference stars.}
\end{figure}

\section{Results.}

We have plotted in Fig.3 the relative positions of our target together
with  the  residuals on  the  reference  stars  positions.  Since  the
difference between  the positions  of Cir X-1  is comparable  with the
residuals measured  for the 42  reference stars, we conclude  that the
Cir X-1  counterpart does not show any  significant displacement.  The
position  offsets   amount  to  0.24$\pm$0.33  PC  pixel   in  RA  and
-0.15$\pm$0.25 PC pixel  in Dec.  By applying the  nominal plate scale
tranformation (PC pixel size 0\farcs0455), these values translate into
an angular displacement  in the plane of the sky  of -11$\pm$15 mas in
RA and -7$\pm$11  mas in Dec (neglecting the  correction for the Sun's
proper motion).   Although such displacements  are statistically null,
it is  worth comparing  the derived proper  motion with  the predicted
values.   Over the  time  interval of  8.6  years spanned  by our  HST
observations, we obtain a total yearly displacement of 1.5$\pm$1.6 mas
yr $^{-1}$,
setting the $3 \sigma$ upper limit on the proper motion to $\approx$ 5
mas yr  $^{-1}$.  Such  a value  rules out even  the minimum  value of
$\approx$ 15  mas yr$^{-1}$ required by the  proposed association with
G321.9-0.3 (see  Sect.  1).  Moreover,  we note that also  the derived
constraints  on the possible  proper motion  directions (Fig.   3) are
 incompatible with the relative  positions of Cir X-1 and the SNR
center, which are roughly aligned  North-South (see Fig.  1). Thus, we
conclude  that, contrary  to  earlier speculations,  Cir  X-1 and  SNR
G321.9-0.3 are not associated.

\section{Discussion.}


Both  the lack  of a  measurable  proper motion  for Cir  X-1 and  its
resulting  non-association   with  SNR  G321.9-0.3   have  interesting
astrophysical implications. Firstly, it  is important to note that our
result frees  the neutron  star age   from the  tight constraints
imposed by the proposed SNR  association.  Indeed, an age greater than
100\,000 years would be more easily compatible with the estimated time
scales  of the  accretion-driven decay  of the  neutron  star magnetic
field  (see, e.g.,  Konar  \& Bhattacharya  1997,  1999 and  reference
therein) down to the values expected for the occurance of type I X-ray
bursts (Lewin  \& Joss 1983).  \\   As far as the  dynamics of the
binary system are concerned,  the null proper motion measurement sets
a $3 \sigma$  upper limit of $24~d_{kpc}$ km  s$^{-1}$ on its relative
tranverse  velocity, where $d_{kpc}$  is the  source distance  in kpc.
Such a value would imply that  Cir X-1 is not moving with an anomalous
velocity  larger  than,  e.g.,  400  km s$^{-1}$  unless  its  spatial
velocity  is essentially radial.   A possible  indication for  an high
radial velocity  ($\approx 430$ km  s$^{-1}$) was indeed found  in the
spectroscopic  observations   of  the  Cir   X-1  optical  counterpart
(Johnston, Fender \& Wu 1999),  although it could not be unambiguously
confirmed by the more recent  work of Johnston et al. (2001). Together
with   our  constraint   on  the   tranverse  velocity,   an  accurate
spectroscopic measurement of the  radial velocity would be critical to
determine the  dynamics of the binary  system and to  compute the kick
velocity  impressed  by  the   SN  explosion  (Tauris  et  al.  1999).
Moreover, the  above  constraints  on the binary  system dynamics
 may require a different scenario for the source interactions with
the  surrounding environment.  In  particular, the  lack of  a proper
motion  in the  expected direction  prompts  for  an alternative
explanation  for  the  morphology  of the  arc-shaped  radio  features
observed  in  the  synchrotron  nebula,  so far  interpreted  as  jets
trailing behind by a fast-moving  source.  \\ Last, but not least, our
result  calls, once  again, for  more caution  in  associating compact
objects with SNRs  purely on  the basis of their proximity in the
plane of the sky and on morphological arguments.

\begin{acknowledgements}
We thank the  referee, H. Johnston, for her  useful  comments.  We are
also   grateful  to G.F. Bignami  for   his  critical  reading  of the
manuscript. 
\end{acknowledgements}



\end{document}